\documentstyle[preprint,aps]{revtex}

\tighten
\begin{document}

\title{
  On the Staggering Effect
  of Dynamical Moments of Inertia in Superdeformed Bands
}

\author     {
             P. Magierski \\
          {\em Institute of Physics, Warsaw University of Technology,}\\
          {\em ul. Koszykowa 75, PL--00-662 Warsaw, Poland } \\
            }

\maketitle
\narrowtext

\begin{abstract}
The possibility of the appearance of the C$_{4}$ symmetry
in the
rotational bands is studied within the particle-rotor
model.  The role of the triaxiality of the rotor is analized.
\end{abstract}


\section{INTRODUCTION}
In the spectroscopy of superdeformed
(SD) bands in $^{149}$Gd \cite{Fli93}, $^{153}$Dy \cite{Ced}, $^{194}$Hg
\cite{Ced94}, and $^{131,132}$Ce \cite{Sem95}
the $\Delta I=4$ staggering of the dynamical moment of inertia
has been observed.
It manifests itself in systematic shifts of the energy levels which are
alternately pushed down and up with respect to a purely rotational sequence.
The amplitude of this staggering is of the order of 50 eV.
It was suggested that their origin could be associated with
the presence of the C$_4$ symmetry.

To date some models have been proposed to
explain the experimental data.
Hamamoto and Mottelson \cite{HM94,HM95} have studied the properties of
a quartic rotational Hamiltonian.
In their approach
the C$_4$ perturbation coincides with the
symmetry axis of a nucleus.
Staggering appears then as a
result of tunneling between the four
equivalent minima of the total energy surface
of the Hamiltonian due to the $K$-mixing.
Nevertheless microscopic approaches do not confirm the presence
of such perturbation in the mean field \cite{Rag94,Fra94,MH95,L95},
or they found it to be too small to generate the effect.

On the other hand Pavlichenkov and Flibotte
 \cite{Pav95} proposed the model in which
C$_4$ perturbation appears along the rotation axis.
They employ the rotational Hamiltonian which generate
the C$_{4v}$ bifurcation. In this scenario staggering effect
does not originate from a static hexadecapole deformation. Rather,
it arises from a dynamical effect that involves the alignment of an
angular momentum vector.

One should also mention that the effect of staggering
has been generated by the hexadecapole-hexadecapole interaction
in the simple model consisting of a single j-shell filled by $N$
identical nucleons \cite{BM94,MB94}.

In the present paper
I consider the triaxial version
of the particle plus rotor model. Allowing for a slight triaxiality
of the rotor it is shown that the lowest states of the
system for a sufficiently high spin possess the symmetry which allows
to distinguish states differing by two units of angular momentum.

\section{THE MODEL}

The model consists of a single particle occupying j-shell and coupled to
the triaxial core in the strong coupling limit:
\begin{equation}\label{ham}
\hat{H} = \sum_{\alpha}\Bigg{[}A_{\alpha}
          (\hat{I}_{\alpha}-\hat{j}_{\alpha})^{2} +
          Q_{\alpha}\hat{j}_{\alpha}^{2}\Bigg{]} +
         \chi (\hat{j}_{1}^{2}-\hat{j}_{2}^{2})^{2} .
\end{equation}
In the above formula $\hat{I}_{\alpha}$ denotes the total angular
momentum component
of the rotor in the body-fixed frame,
whereas $\hat{j}_{\alpha}$ stands for the
single-particle spin component. The coefficients $A_{\alpha}$ are inversely
proportional to moments of inertia according to the relation
$\displaystyle{A_{\alpha}=\frac{1}{2{\cal J}_{\alpha}}}$.
The $Q_{\alpha}$ are the mean-field parameters describing
the quadrupole deformation of the shell whereas $\chi$ is the strength
of the C$_{4}$ type of deformation along the third axis\footnote{
This model has been first studied (without the last term in (\ref{ham}))
by Pashkevich and Sardaryan \cite{PS65}.}.
In the further considerations I will assume that $A_{2}>A_{1}\ge A_{3}$.
The diagonalization of the above Hamiltonian
is performed in the basis of states:
\begin{equation}\label{basis}
|IMKjk\rangle = |IMK\rangle |jk\rangle ,
\end{equation}
where $|IMK\rangle$ denotes the  state of the rotor
and can be expressed as Wigner functions depending
on Euler angles. By $M$ I denoted the projection of
the total angular momentum $I$ on the third axis in the
laboratory frame whereas $K$ denotes the projection
on the third axis in the body-fixed frame.
Since the Hamiltonian is rotationally invariant I will omit
the quantum number $M$ in the formula (\ref{basis}).
The  $|jk\rangle$  describes
the single particle state with spin $j$ and projection
on the third axis in the body-fixed frame equal to $k$.

If one assumes that the particle is coupled to a core
rotational band comprising spins $0, 2, 4, ...$ then the
basis will be restricted to states for which the following relation
holds:
\begin{equation}
|K-k| = 0, 2, 4, ...
\end{equation}
Hence the result eigenfunctions of (\ref{ham}) corresponding
to the total spin $I$ will have the form
\begin{equation}
|\Psi_{I}\rangle = \sum_{K,k} a_{K,k} |IjKk\rangle .
\end{equation}
The classical motion of the vectors ${\bf j}$
and ${\bf I}$ can be determined from the following relations:
\begin{equation}
\begin{array}{cll}
\dot{j}_{i}&=&\{j_{i}, H\} \\
\dot{I}_{i}&=&\{I_{i}, H\}, \\
\end{array}
\end{equation}
where $\{ . , . \}$ denotes the Poisson bracket.
Thus the time derivatives of the components of ${\bf I}$
and ${\bf j}$
follow from the transformation of the Hamiltonian under infinitesimal
rotations.
The qualitative nature of the motion corresponds to a periodic
precession of ${\bf I}$ and ${\bf j}$ around the minimum
for sufficiently low energies. If the energy becomes high enough
the structure of orbits will change dramatically.
This critical value of energy
defines the separatrix on the total energy surface which
divides the phase space into separated regions\footnote{
It has been studied in case of an axially symmetric
rotor by Bohr and Mottelson \cite{BM80}.}.

In order to investigate the small amplitude
motion near the minimum point
one can assume that
the angular momentum of the core ${\bf R}={\bf I} - {\bf j}$
is aligned along the third axis. This is the case when
$A_{3} < A_{1,2}$. Thus one can put
$I_{2}-j_{2}\approx 0$ and $I_{1}-j_{1}\approx 0$.
The above equations can be then simplified  to the form:
\begin{equation}
\begin{array}{cll}\label{orbit1}
\dot{j}_{1}&=&
2A_{3}j_{2}(I_{3}-j_{3}) + 2j_{2}j_{3}(Q_{2}-Q_{3})-
4\chi j_{2}j_{3}(j_{1}^{2}-j_{2}^{2}) \\
\dot{j}_{2}&=&
-2A_{3}j_{1}(I_{3}-j_{3}) - 2j_{1}j_{3}(Q_{1}-Q_{3})-
4\chi j_{1}j_{3}(j_{1}^{2}-j_{2}^{2}) \\
\dot{j}_{3}&=&
2j_{1}j_{2}(Q_{1}-Q_{2})+
8\chi j_{1}j_{2}(j_{1}^{2}-j_{2}^{2}) \\
\dot{I}_{1}&=&-2A_{3}I_{2}(I_{3}-j_{3}) \\
\dot{I}_{2}&=&2A_{3}I_{1}(I_{3}-j_{3}) \\
\dot{I}_{3}&=&0 \\
\end{array}
\end{equation}

One can see that in case of $Q_{1}=Q_{2}$ the equations
involving vector ${\bf j}$ possess the symmetry associated
with the rotation around the third axis about the $\frac{\pi}{2}$
angle (i.e. C$_{4}$ symmetry).
That means that the transformation
\begin{equation}
\left\{
\begin{array}{cccc}\label{c4}
j_{1}\rightarrow j_{2} \\
j_{2}\rightarrow -j_{1} \\
I_{1}\rightarrow I_{2} \\
I_{2}\rightarrow -I_{1} \\
\end{array}
\right .
\end{equation}
leaves the equations (\ref{orbit1}) unchanged.
Therefore the orbits around the minimum will possess the additional
symmetry originating from the local properties
of the total energy surface in the vicinity of the minimum point.
In our case the transformations (\ref{c4}) form the local symmetry group
which is suitable for the description
of a part of the rotational multiplet levels.
Obviously such symmetry will be satisfied only approximately and
the amplitude of the symmetry breaking components will depend
on the relative magnitude of symmetry breaking and restoring terms
in (\ref{orbit1}) as well as the energy of the motion.

\section{RESULTS AND CONCLUSIONS}

In the current section I will present the results of
quantal calculations performed within the model described in the
previous section. The diagonalization has been performed
for a given spin $I$ with one particle occupying the level
$j=\displaystyle{\frac{11}{2}}$.
In order to estimate whether the symmetry
is broken completely or it survives one can introduce the
measure of C$_{4}$ symmetry possessed by a given state.
It is defined as the absolute value of a difference between
the components of a wave function associated with different
representations of this group. Thus for the wave function
\begin{equation}
|\Psi_{I}\rangle = a |\Psi_{I}^{1}\rangle + b |\Psi_{I}^{2}\rangle
\end{equation}
the quantity $||a|^{2}-|b|^{2}|$
will be the measure of the C$_{4}$ symmetry.
As an example one can
consider Fig. 1 where this quantity is plotted
as a function of the total angular momentum.
Calculations were performed for $A_{3}= 10 keV$, $A_{1}=21 keV$,
$A_{2}=240 keV$ for three different sets of parameters. One can
see that the shape of curves looks similarly.
As the spin increases the motion of the system approaches the
aligned regime discussed in the previous section.
In spite of the fact that the Hamiltonian does not possess the C$_{4}$
symmetry the lowest levels can be characterized by a suitable
quantum number associated with this symmetry. Moreover one should
emphasize that the quantum number associated with the C$_{4}$ symmetry
will change when the spin increases. This can
be easily understood if one considers the lowest state of the Hamiltonian
to be approximately
\begin{equation}
|\psi_{I}\rangle \approx |IjIj\rangle
\end{equation}
which corresponds to the aligned configuration. One can see
that $|\psi_{I}\rangle $ and $|\psi_{I+2}\rangle $
belong to different representations of the C$_{4}$ group (this
fact has been pointed out in \cite{Pav95}).
Moreover one need not necessarily employ the C$_{4}$ term ($\chi\ne 0$)
to generate this symmetry.

It is obvious  that the symmetry can manifest itself only
in the case when we allow for a small triaxiality of the rotor.
This is shown in Fig. 2 where the
quantity $||a|^{2}-|b|^{2}|$
for the lowest states
is plotted versus
$\displaystyle{ \frac{{\cal J}_{1}}{{\cal J}_{3}} }$. Calculations
have been performed for the same sets of parametrs as before.
One can see that for an axial rotor the symmetry is completely
broken. This is the consequence of the fact that in this case
the spin of the core ${\bf R}$ is delocalized and therefore breaks
the rotational symmetry around the third axis.

This research was supported in part by the Polish State Committee for
Scientific Research under Contracts No. 2 P302 056 06
and 2 P30 B 09809, and by the
computational grant from the Interdisciplinary Centre for Mathematical
and Computational Modelling (ICM) of Warsaw University.

\newpage

\begin{figure}[ht]
\caption[FIG2]{%
The quantity $\displaystyle{||a|^{2}-|b|^{2}|}$  plotted
for the lowest states
versus total angular momentum for three different sets of parameters.
$Q_{2}=200$, $Q_{3}=0$, $A_{3}= 10 $, $A_{1}=21 $ , $A_{2}=240$ keV
}
\label{FIG2}
\end{figure}

\begin{figure}[ht]
\caption[FIG3]{%
The quantity $\displaystyle{||a|^{2}-|b|^{2}|}$  plotted
versus $\frac{J_{1}}{J_{3}}$ for the same
sets of parameters as in the figure 1 and $I = \frac{101}{2}$.
}
\label{FIG3}
\end{figure}

\end{document}